\documentclass[preprint,showpacs,amsfonts,aps,prl,superscriptaddress]{revtex4}

\usepackage{amsmath}
\usepackage{graphicx}
\usepackage[latin1]{inputenc}
\usepackage{times}

\begin{document}
\title{Motion by Stopping: Rectifying Brownian Motion of Non-spherical Particles}
 \date{\today}
\author{Susan Sporer}

\author{Christian Goll}

\author{Klaus Mecke}
%  \email{Klaus.Mecke@physik.uni-erlangen.de}

\affiliation{Institut f\"ur Theoretische Physik, 
Universit\"at Erlangen-N\"urnberg, Staudtstra\ss e 7, 91058 Erlangen, Germany}

\begin{abstract}
We show that Brownian motion is spatially  not symmetric for mesoscopic particles embedded in a fluid 
 if the particle is not in thermal equilibrium and its shape is not spherical. 
In view of applications on molecular motors in biological cells, we sustain non-equilibrium 
by stopping a non-spherical particle at periodic sites along a filament.
Molecular dynamics simulations in a Lennard-Jones fluid demonstrate that directed motion is possible 
without a ratchet potential or temperature gradients if the asymmetric non-equilibrium relaxation 
process is hindered by external stopping. Analytic calculations in the ideal gas limit show that 
motion even against a fluid drift is possible and that the direction of motion can be controlled 
by the shape of the particle, which is completely characterized by tensorial Minkowski functionals. 
\end{abstract}


%   05.70.Ln: Nonequilibrium and irreversible thermodynamics 
%   05.20.Dd: Kinetic theory
%   05.40.Jc: Brownian motion
%   05.60.Cd: Classical transport 
%   87.15.Vv: Diffusion  (of Biomolecules)  
%   87.16.Ac: Theory and modeling; computer simulation  (for subcellular processes) 

\maketitle

Mesoscopic particles dissolved in a fluid are expected to perform symmetric thermal 
fluctuations around a mean position \cite{einstein05,smoluchowski06}. 
The shape of the particles is irrelevant for this so-called Brownian motion - 
as long as the particle is in thermal equilibrium with the fluid. 
Consequently, a net transport in a preferred direction is not possible without 
applying an external force which breaks the spatial symmetry. 
In the past, several models for Brownian motors have been proposed 
 based on different methods of rectifying thermal noise; 
for instance, by an  asymmetric external potential  which is switched on and off periodically
 (for a review see Ref.~\cite{reimann:2002}). 
Another concept is  based on two heatbaths at different temperatures such as the 
Feynman ratchet \cite{feynman:63} and its simplification by Van den 
Broeck \cite{broeck-meurs-kawai:2004}. 
It was shown that  directed Brownian motion can be achieved as long as the two reservoirs have different temperatures and spatial symmetry is broken in some way. 
This letter demonstrates that  a Brownian motor can be built even in a single 
heatbath without violating the second law of thermodynamics. Hence, it is  a possible 
theoretical model for molecular motors in biological  cells.  
We show that a steady state motion of non-spherical particles is  obtained 
if their relaxation  towards equilibrium is prohibited by  periodical stopping.

To illustrate the concept we restrict 
the system  to two dimensions and follow  
Ref.~\cite{broeck-meurs-kawai:2004}. Since the spatial 
symmetry breaking is a prerequisite for net transport, we
consider an asymmetric 
motor $K$ of mass $M$ in a single heatbath at temperature T. 
The motion of the motor is restricted to a one-dimensional track, here the $x$-axis, so 
that the motor velocity can be written as ${\bf \widetilde{V}}=(V\sqrt{k_BT/M},0)$ with 
the normalized velocity $V$ in $x$-direction.
Furthermore, elastic interactions are assumed  between the motor $K$ and the surrounding fluid 
particles. 
In contrast to Refs.~\cite{cleuren:2007,costantini:2007} there is no dissipation involved 
in this collision.  
However, due to the  confinement only the axial momentum component is conserved.
In thermal equilibrium, directed motion in a single heatbath is prohibited by the 
second law of thermodynamics. The motor $K$  behaves like a Brownian particle 
with a Maxwellian velocity distribution with mean $\langle V \rangle =0$ 
and variance 
$\langle V^2 \rangle = 1$; 
independent of its particular shape. Hence, in order  to get directed motion, we have 
to sustain a non-equilibrium state.  

Our model is motivated by a molecular motor in a cell, e.g.  kinesin, that binds 
to a filament after every step \cite{howard:1996}. 
Similarly, our particle is stopped at periodically spaced binding sites along the track, 
i.e. its kinetic energy is set to zero whenever it reaches a binding site. Thermodynamically, 
of course, the stopping requires a greater amount of work  than the decrease in entropy; 
here ${1\over 2} k_BT$ from the reduction of the Maxwellian distribution for $V$ to 
the velocity distribution $P(V)=\delta (V)$ for the stopped particle.  
After binding the motor is released again and starts to relax  towards thermal equilibrium. 
By means of collisions between fluid particles and the motor, kinetic energy is transfered 
from the fluid to the motor until the motor has the same effective temperature as 
the surrounding bath.

\begin{figure}[b]
\includegraphics[width=0.49\linewidth]{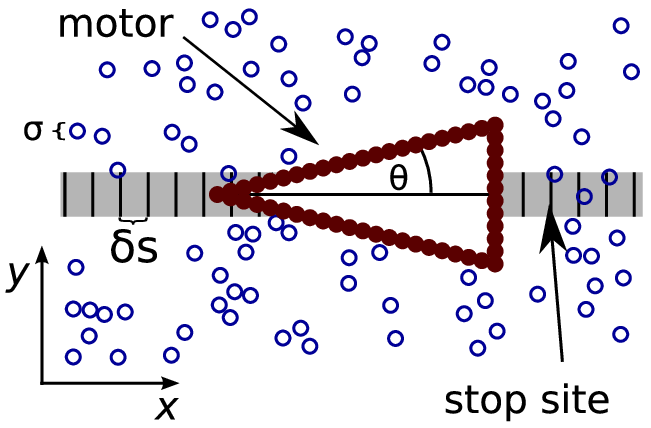}
\hfill
\includegraphics[width=0.49\linewidth]{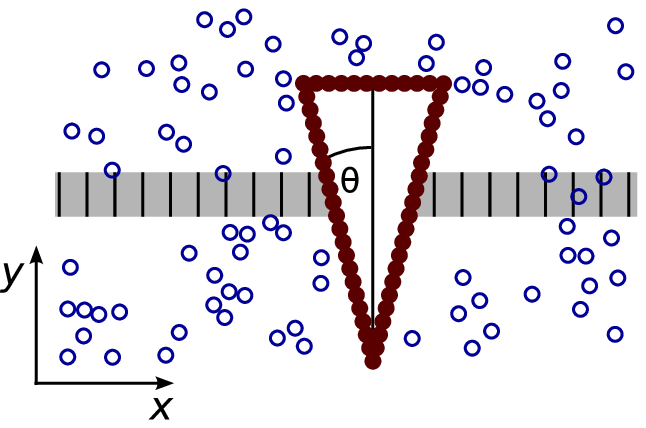}
\caption{An asymmetrically shaped motor (here a triangle) is built from fluid particles and 
placed in a Lennard-Jones fluid. Its motion is restricted to a one-dimensional track with 
periodically spaced binding sites along the x-axis. If the motor's center of mass 
crosses a binding site, 
the velocity of the motor is set to zero. To test the importance of the motor's shape for the 
momentum transfer during collisions with fluid particles  we orient the triangle in two 
different ways. }
\label{fig:sim_fig}
\end{figure}

In the following, we show that this non-equilibrium relaxation process is asymmetric due 
to the asymmetry of the particle. This asymmetric relaxation process  can be used to 
rectify thermal fluctuations.  
To test this hypothesis, we performed 2D molecular dynamics simulations  of an asymmetric particle 
in a two-dimensional fluid. The interaction between the fluid particles is described 
by the Lennard-Jones (LJ) potential 
$\mathcal{U}(r)= 4 \mathcal{E} \left( \left( \frac{\sigma}{r} \right)^{12} 
- \left( \frac{\sigma}{r}\right)^6 \right)$ 
where $r$ is the distance between two particles with diameter  $\sigma$ and $\mathcal{E}$ is 
the depth of the potential well. 
In the following, all lengths are given in units of $\sigma$, energies in units of $\mathcal{E}$ 
and times are expressed in units of $t_0=\sigma \sqrt{m/\mathcal{E}}$ where $m$ denotes the mass 
of the fluid particles. 
The simulation box with edge length $100 \sigma$ contains  $n=800$ fluid particles yielding a 
particle density  $\varrho \approx 0.08 \sigma^{-2}$. The temperature is set to 
$T=3 \mathcal{E}/k_B$ and the overall simulation time was $2000 t_0$. With these parameters 
the LJ fluid is in the gas phase without any long range correlations. Periodic boundary conditions  
are  used  and also a cut-off radius of $2.5 \sigma$ in $\mathcal{U}(r)$ 
to limit the range of the potential. 

As illustrated in Fig.~\ref{fig:sim_fig} the motor is built of $N$ stiffly-linked fluid particles 
which 
interact  with the surrounding fluid also by a Lennard-Jones potential. 
Consequently, the motor mass $M=Nm$ scales with the mass $m$ of the fluid particles. 
The motion of the motor is confined to a one-dimensional track with periodically spaced 
stopping sites. Every time the motor's center of mass reaches one of those sites, 
the velocity of the motor is set to zero and the corresponding kinetic energy is 
re-transferred to the fluid.

\begin{figure}
\includegraphics[width=0.8\linewidth]{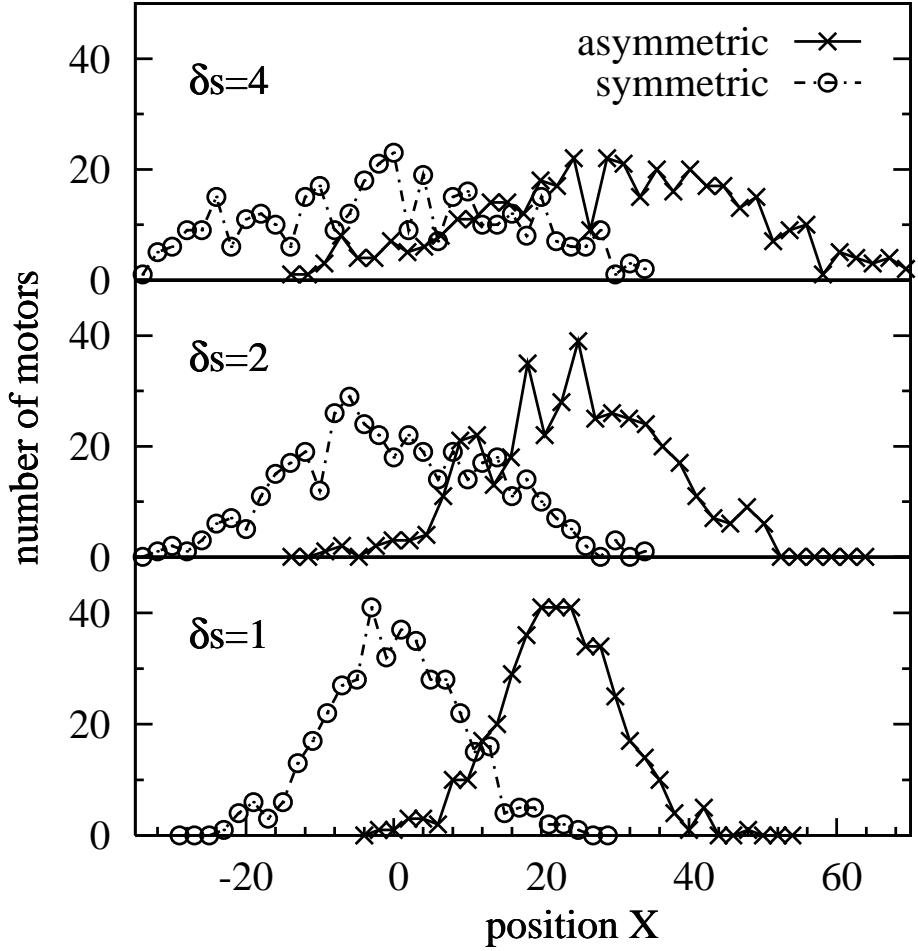}
\caption{Histograms  of end positions at time $t=2000t_0$ 
of a triangular motor where the distance between stopping sites is 
$\delta s$ (see Fig.~\protect\ref{fig:sim_fig}). The asymmetrically oriented motor moves  preferable 
to positive X-values whereas the symmetric triangle diffuses around $X=0$. 
A discrete spacing is visible since the motors accumulate around the stopping sites. 
400 realisations are used with integration step $\delta t=0.001 t_0$ and $2000000$ steps.
With decreasing  distances $\delta s$ between stopping 
sites the distribution becomes narrower, as expected for a hindered diffusion process.
}
\label{fig:endpositions}
\end{figure}

\begin{figure}
\includegraphics[width=0.9\linewidth]{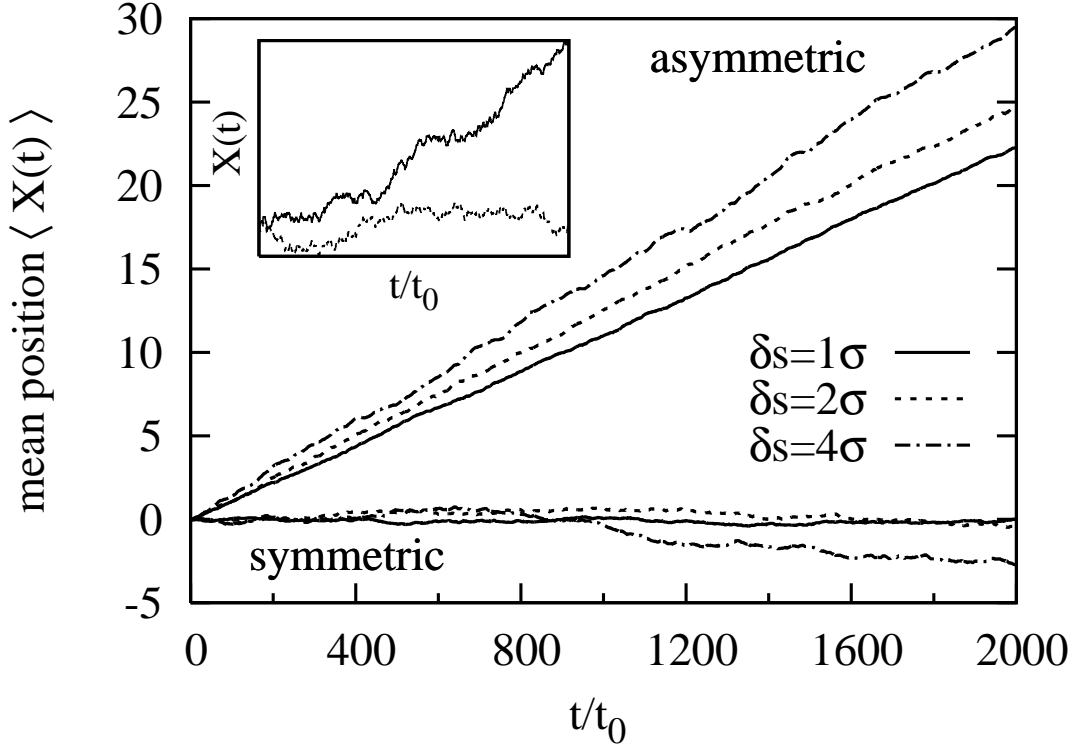}
\caption{The average  position (400 runs)  of an asymmetrically oriented, 
isosceles triangle of height $10\sigma$ and basis length $20\sigma$ moves 
with a constant velocity in the direction  opposite to its sharp edge 
(i.e., in the direction of its  basis). 
Thus, directed motion is possible 
when the triangle is oriented  asymmetrically 
with respect to the direction of motion. Otherwise the motor 
performs symmetric fluctuations around its starting position.  
Inset: trajectories of individual motors. 
Note, that the 
velocity decreases for decreasing distances $\delta s$ between stopping 
sites if the typical stopping time is smaller than the relaxation time of the motor. 
}
\label{fig:sim_results}
\end{figure}

Consider a motor with the shape of an isosceles triangle  of height $10\sigma$ 
and basis length $20\sigma$ which corresponds to an apex angle of $\theta=0.25268$, $N=53$ 
and  mass $M=53 m$. The motor is placed in the beginning ($t=0$) at the position $X=0$. 
Figs.~\ref{fig:endpositions} and \ref{fig:sim_results} show the simulation results for 
the motion of the motor. 
Note, that for the chosen parameters 
the typical time between two successive stops of the motor is smaller than the relaxation 
time into equilibrium.  
In case of an isosceles triangle, the motor can be 
oriented symmetrically or asymmetrically with respect to its direction of motion 
(Fig.~\ref{fig:sim_fig}), to test the importance of the shape.  
Fig.~\ref{fig:endpositions} presents the distribution over multiple runs of the end positions, 
i.e. the positions of 
the motors after time $2000t_0$, for different stopping distances $\delta s$. 
If the triangle is placed with the apex onto the track, i.e. asymmetric with respect to the 
direction of motion, the maximum  of the distribution of the end positions is shifted to 
positive values $x>0$. Hence, on average, the triangle travels a certain distance towards 
its basis. This is in accordance with the results shown in Fig.~\ref{fig:sim_results} where a 
net speed of about $\langle v \rangle = 0.011 \frac{\sigma}{t_0}$ in the direction of the 
triangle basis can be observed for asymmetric orientation. 
In contrast, aligning the triangle basis  parallel to the $x$-axis conserves spatial 
symmetry, i.e. mirror-symmetry with respect to the y-axis, and therefore  
 causes the mean position to fluctuate around zero without any net motion 
(see Fig.~\ref{fig:sim_results}). The end positions of the motors are in this case 
symmetrically distributed around the origin (see Fig.~\ref{fig:endpositions}). 
This shows two things: first, the proposed model demonstrates the feasibility of rectifying 
Brownian motion in a single heatbath by stopping the particle at periodic sites.  
Second, it is sufficient to break the spatial symmetry by the shape of the particle, so that 
an asymmetric external potential is not  necessary.

To get a better insight into the microscopic mechanism of the driving force, we  calculate 
analytically   the stochastic dynamics of this non-equilibrium relaxation process for a simplified model. 
The  interaction between the fluid and the motor   is no longer a 
Lennard-Jones potential but an ideal hard sphere interaction with infinitesimal 
interaction time. 
The total energy is conserved as well as the momentum in $x$-direction and tangential to 
the particle's surface. Details of this model 
can be found in  Ref.~\cite{broeck-meurs-kawai:2004}. 
 
Furthermore, we assume  an  infinite heatbath,  so that the temperature is constant 
and the velocity distribution of the fluid is a Maxwell distribution 
$\Phi_{\boldsymbol u}({\boldsymbol v})=\frac{m}{2 \pi k_B T} \exp 
\left( - \frac{m}{2 k_B T} ({\boldsymbol v}-{\boldsymbol u})^2 \right)\, $. 
To study motion against a drift we introduce  an average drift  velocity  $\boldsymbol u$ 
of the fluid. In the beginning, we assume ${\boldsymbol u}=0$. In order to exclude multiple collisions, 
we use the ideal gas limit of the fluid 
and confine the motor shape to convex objects $K$  so that 
the stochastic motion is a Markov process.  
This allows us to describe the time dependence of the velocity distribution 
$P(V,t)$ of the motor by a Master equation with a 
transition rate \cite{broeck-meurs-kawai:2004}
\begin{align}
\label{eqn:transitionprob}
W(V'|V)&= \int_{\partial K} d\mathcal{O} \int_{-{\infty}}^{+\infty} d{\boldsymbol v} 
\; \rho \; \Phi_0({\boldsymbol v}) \, ({\bf \widetilde{V}}-{\boldsymbol v}) \cdot {\bf \hat{e}}_{\perp}  \nonumber \\
& \times \Theta \big[ ({\bf \widetilde{V}}-{\boldsymbol v}) \cdot {\bf \hat{e}}_{\perp} \big] 
\;  \delta \big[ \widetilde{V}'-\widetilde{V} + \frac{\Delta\boldsymbol p}{M}  \cdot {\bf \hat{e}}_{x} \big] \, 
\end{align}
($\widetilde{V}=\sqrt{k_BT/M} \, V$) which equals the number of collision processes from $V$ to $V'$ 
per unit time 
using the dimensionless  time 
$t =  8  \epsilon ^2  \sqrt{\frac{k_B  T  \rho}{2  \pi  m}}  \tilde{t}$.   
Here, ${\bf \hat{e}}_{\parallel}=(\cos \phi, \sin \phi)$ is the tangential and 
${\bf \hat{e}}_{\perp}=\left( \sin \phi, - \cos \phi \right)$ the normal vector in the 
collision point ${\bf r} \in \partial K$ on the boundary $\partial K$ of the motor  
and $\phi$ is the polar angle.
The momentum transfer of a single collision is then given by 
$\Delta{\boldsymbol p}=\frac{2 \; m\, M{\bf \hat{e}}_{\perp}}{M + m \left( {\bf \hat{e}}_{\perp} \cdot  
{\bf \hat{e}}_{x} \right)^2} \,\left( {\bf \widetilde{V}}-{\boldsymbol v} \right) \cdot  {\bf \hat{e}}_{\perp}$ 
 \ (see Ref.~\cite{broeck-meurs-kawai:2004}).  
Because the Master equation 
cannot be solved exactly, we  apply 
a Kramers-Moyal expansion yielding 
\begin{align}
\label{eqn:kramers-moyal}
\frac{\partial P (V,t)}{\partial t} \;
& = \; \sum_{n=1}^{\infty} \frac{1}{n!} \left( - \frac{\partial}{\partial V} \right)^n  
\big[ A_n(V,t) P(V,t) \big] 
\end{align}
with the coefficients 
$A_n(V,t) = \int dV' (V'-V)^n W(V|V')$. 
Different to Ref.~\cite{broeck-meurs-kawai:2004}, we emphasise two important points: 
(i) the dependence on  the shape of the motor can be fully  described  by tensorial 
Minkowski functionals $M_\nu^{(r,s)}(K)$ of the motor $K$;  
(ii)  the non-stationary solution of Eq.~(\ref{eqn:kramers-moyal}) is the essential 
ingredient to rectify Brownian motion if relaxation is 
prohibited by stopping sites    
(for details see Ref.~\cite{sporer:2006}).

Tensorial Minkowski functionals are defined as  surface integrals  
in $d$ dimensions \cite{lecturenotes02}  
\begin{align}
\label{eqn:def_minkowski_tensor}
M_1&^{(r,s)}(K) =  \int_{\partial K} \!\!\! d\mathcal{O} \;  \overbrace{{\bf r} \otimes \ldots \otimes {\bf r}}^{r-times} \otimes \overbrace{{\bf \hat{e}}_{\perp} \otimes \ldots \otimes {\bf \hat{e}}_{\perp} }^{s-times} 
\end{align} 
over tensor products of  position  ${\bf r}$ and normal vector ${\bf \hat{e}}_{\perp}$ 
on the surface $\partial K$ of motor $K$.
Using 
$ \left( {\bf \hat{e}}_{\perp} \cdot { \bf \hat{e}}_{x} \right)^n= \sin^n \phi$,  
the $x \ldots x$-components of the first tensorial Minkowski functional in two dimensions 
can be written as 
$\left( M_1^{(0,n)}(K) \right) _ {\overbrace{x \ldots x}}^{n-times}  
=  \int_{\partial K} d\mathcal{O}  \sin^n \phi$.  
For convex bodies $K$ one finds $M_{\nu}^{(0,1)}(K) =  0$ 
so that  the first two jump moments read  
$A_1(V)=\tau^{-1}(V+4V_{\rm max}(V^2-1))$ and  
$A_2(V)=2\tau^{-1}(1+12V_{\rm max}V)$. 
Here, we introduced  the relaxation rate 
$\tau^{-1}=\sqrt{\rho} \left( M_1^{(0,2)} (K) \right)_{xx}/2$ and  
the maximum velocity  
\begin{equation}
\label{maxvel} 
V_\text{max}={\epsilon \over 4} \, \sqrt{\frac{\pi}{8}} 
\frac{\left(M_1^{(0,3)}(K) \right)_{xxx}}{\left(M_1^{(0,2)}(K) \right)_{xx}}  \;\;. 
\end{equation} 
The shape dependence of the stochastic dynamics enters only via these two 
parameters which can be 
expressed in terms of Minkowski tensors. These functionals are known in many cases 
to allow for a complete description of the shape-dependence of physical properties \cite{lecturenotes02}. 
The other Minkowski tensors in Eq.~(\ref{eqn:def_minkowski_tensor}) play  a role when the collision 
rules in Eq.~(\ref{eqn:transitionprob}) are modified or the drift is non-zero, for instance \cite{sporer:2006}. 

From Eq.~(\ref{eqn:kramers-moyal}), one  immediately obtains the time dependence of the 
average velocity 
$\partial_{t} \langle V \rangle = \langle A_1(V) \rangle$ 
and the average squared velocity 
$\partial_{t} \langle V^2 \rangle 
= 2 \langle V A_1(V) \rangle + \langle A_2(V)\rangle$.  
The equations are coupled to higher moments of $V$ because of the dependence of 
$A_n(V)$  on  $V$. 
However, assuming the mass $m$ of the fluid particles 
to be small against the motor mass $M$, an expansion in $\epsilon=\sqrt{\frac{m}{M}}$ 
leads to a decoupling of the differential equations  from higher moments when terms 
of order $\mathcal{O}(\epsilon ^4)$ or higher are neglected 
\begin{align}
\label{eqn:dgl_x}
\frac{\partial \langle V \rangle }{\partial t}  &=-  {1\over \tau} \left[\langle V \rangle + \bigg(\langle V^2 \rangle -1 \bigg)V_\text{max} \right] +  \mathcal{O} \left( \epsilon^4 \right) 
\nonumber   \\
\frac{\partial \langle V^2 \rangle }{\partial t} &=  {2\over \tau} 
 \left(1-\langle V^2 \rangle \right)  +  \mathcal{O} \left( \epsilon^4 \right) \; .
\end{align} 
Hence, 
the solutions give readily  the mean velocity 
$\langle V \rangle(t) = 4 V_\text{max} \left( e^{ - t/\tau}  -e^{ -2 t/\tau} \right)$, 
$\langle V^2 \rangle(t) = 1  -e^{ -2 t/\tau}$ 
 and the average of the travelled distance 
$\langle X\rangle(t)=\int_0^{t} dt' \, V(t') = 2 V_\text{max} \tau \left( 1 - e^{-t/\tau}\right)^2$. 
The time $\tau$ characterizes the  increase of the kinetic energy 
${M\over 2} \langle \widetilde{V}^2 \rangle$ towards its equilibrium 
value ${1\over 2} k_BT$, 
which is the stationary solution of Eq.~(\ref{eqn:dgl_x}). 
The relaxation process  depends  on the shape of the motor only via  the maximum of the 
averaged velocity $V_{\rm max}$, so that   
the shape dependence is fully described by the Minkowski functionals 
of the particle $K$. For instance, the Minkowski functionals of a 
triangular motor with the top pointing 
in negative $x$-direction reads 
$(M_1^{(0,2)}(K))_{xx}= L\left(1+ \sin \theta \right)$ and 
$(M_1^{(0,3)}(K))_{xxx} = L\cos ^2 \theta$, so that  
 $V_\text{max}= {\epsilon \over 4} \sqrt{\frac{\pi}{8}} \left(1-\sin \theta \right)$ and 
$\tau^{-1}= L \sqrt{\rho}(1+\sin \theta)/2$ depend on the apex angle $\theta$  
of the triangle (see Fig.~\ref{fig:sim_fig}).

\begin{figure}
\parbox{0.5\linewidth}{
\hspace*{-0.2cm}
\includegraphics[height=7.8cm]{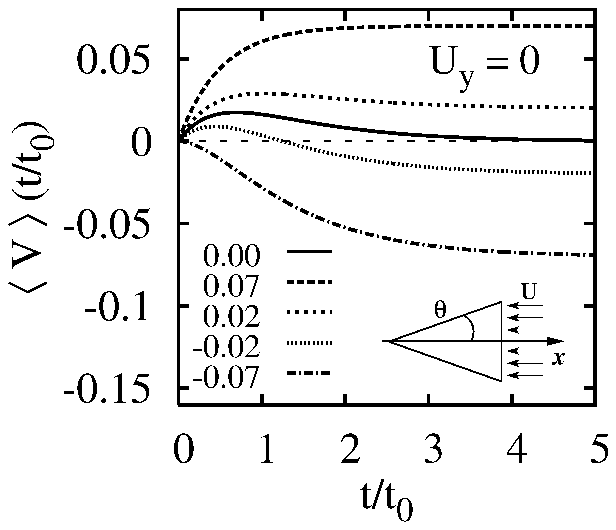}
}
\hfill
\parbox{0.45\linewidth}{
\includegraphics[height=7.4cm]{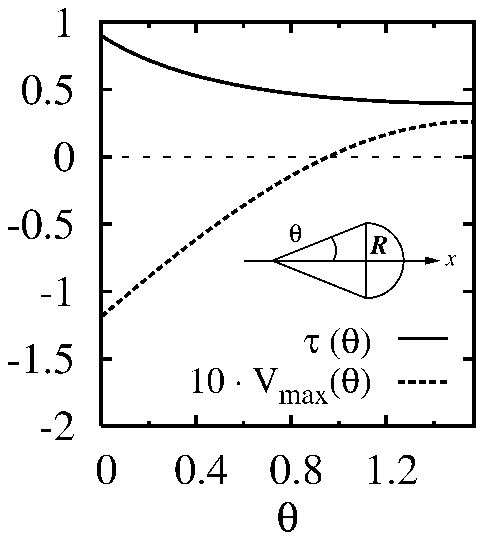}
}
\caption{(a) A triangular motor (apex angle $\theta=\pi/6$, base 
length $L\sqrt{\rho}=1$, $\epsilon ^2 =m/M=0.05$) can swim even against a 
fluid drift velocity $\boldsymbol U = \sqrt{M/(k_B T)} \, {\boldsymbol u}$ 
during its relaxation in thermal equilibrium.  
(b) The direction of motion can even be inverted solely by 
changing the shape of the motor. The maximum velocity $V_{\rm max}$ and  
the relaxation time $\tau$ of a 'capped-triangular'  motor (sketched in 
the inset) 
can be regulated 
by the opening angle $\theta$ which determines the Minkowski functionals 
in Eqs.~(\protect\ref{eqn:def_minkowski_tensor}) and (\protect\ref{maxvel}). }
\label{fig:relaxation_drift}
\end{figure}

The shape dependence of this non-equilibrium Brownian motion can be 
perfectly demonstrated with 
the 'capped-triangular' motor shown in the inset of 
Fig.~\ref{fig:relaxation_drift}(b), whose Minkowski functionals are 
$(M_1^{(0,2)} (K))_{xx}= L\left(\frac{\pi}{4} +  \sin \theta \right)$ and 
$(M_1^{(0,3)} (K))_{xxx}= L\left( \frac{2}{3} -  \; \sin ^2 \theta \right)$.   
As shown in Fig.~\ref{fig:relaxation_drift}(b), 
the relaxation time $\tau$ decreases with increasing opening angle $\theta$ 
whereas the maximum average velocity $V_\text{max}$  
increases with $\theta$ from negative values to positive ones. 
Interestingly, one can change the direction of motion just by adjusting 
the shape of the motor. 
For small $\theta$, i.e. long triangle legs relative to the arc length 
of the semi-circle, the motor moves towards the semi-circle, whereas it 
travels towards the triangle if  $\theta$ is close to $\frac{\pi}{2}$.  
Note, that there is exactly one value for $\theta$ for which the 
motor does not move on average ($V_\text{max}=0$ vanishes) 
though the motor definitely breaks spatial symmetry.

The relaxation process of the triangular motor 
is shown in Fig.~\ref{fig:relaxation_drift}. The average velocity first increases 
until a maximum value $V_\text{max}$ is reached and than decays exponentially 
towards the equilibrium with vanishing mean velocity.
Note, that in the beginning the motor travels an average 
distance $X_{\rm max}=2V_\text{max}\tau $ before it fluctuates around the mean position 
$X_{\rm max}$  in the long time limit. 
Shortly after releasing, the motion of the motor has a preferred direction. 
Thus repeated stopping yields directed motion if it occurs before the system 
has relaxed completely.

A simple analogy of the rectification mechanism is a piston in a cylinder 
separating two fluids in thermal equilibrium at the same temperature. 
If the piston is arrested or has an infinite mass, one finds in thermodynamic 
textbooks that the averaged momentum transfer on one side per unit area 
of its cylindrical cross section equals the pressure of the fluid on 
this side. 
Since the fluids on both sides are equilibrated at the same temperature 
there is no net force 
on the piston. The same is true if the piston can move freely and is in 
thermal equilibrium with the fluids on both sides. However, a piston with 
finite mass $M$ and not in equilibrium experiences kinematic effects 
due to momentum conservation 
 when momentum is transfered by collisions, 
so that the net force  depends on the shape of the piston.

The motor is also able to move against small fluid drifts. 
Repeating the analytic calculations with $\boldsymbol u \neq 0$ 
and confining the drift to the $x$-direction we find the mean velocity 
shown in Fig.~\ref{fig:relaxation_drift}(a) for a triangular motor
 with its apex pointing in negative $x$-direction. 
In the long time limit the motor equilibrates always to the 
drift velocity  
independent of sign and strength 
of the drift. For small drifts in the preferred direction - 
here the positive $x$-direction - the motor accelerates 
to a maximum average velocity $V_\text{max}$ above 
the equilibrium value before it relaxes to the drift velocity $\boldsymbol u$. 
When applying a small drift in opposite direction, it can be seen 
that the motor starts to move in positive direction but changes its 
direction of motion after a certain time. When stopping it repeatedly 
before the direction is changed, an average motion against a fluid drift 
is achieved.

\begin{figure}
\parbox{\linewidth}
{
	\includegraphics[width=0.8\linewidth]{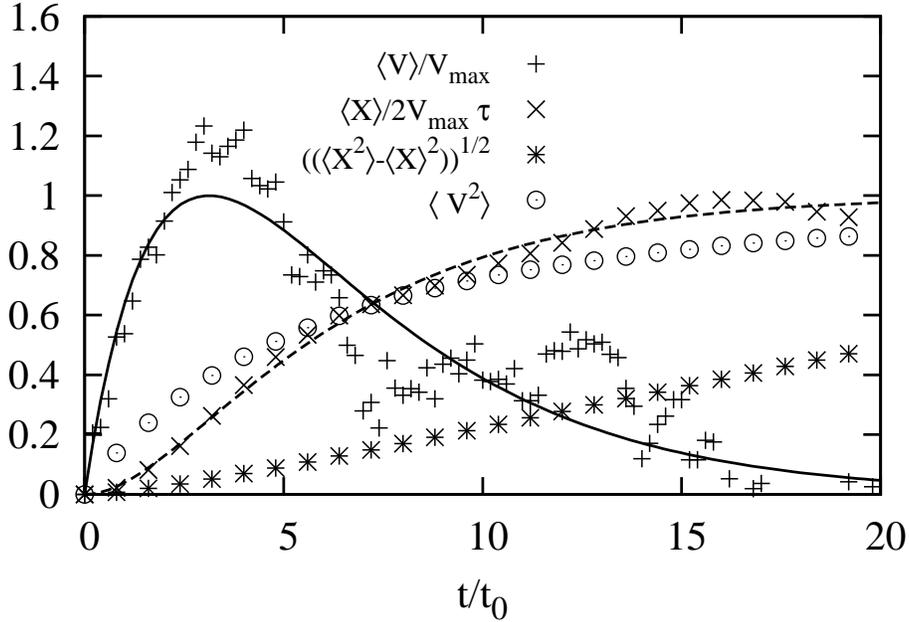}
}
\caption{Relaxation of a motor to thermal equilibrium:  
the molecular dynamics simulations of the LJ-fluid (symbols; averages over 400000 runs) 
can be well described by the analytic results in the  ideal gas limit  (lines)  
if the data are normalized by the maximum velocity $V_{\rm max}$ and the relaxation time. 
}
\label{fig:DataVsTheory}
\end{figure}

Fig.~\ref{fig:DataVsTheory} compares 
our analytic results with the numerical data from the molecular dynamics simulations.  
Evidently, a real fluid with Lennard-Jones interactions cannot be 
in quantitative agreement with calculations based on an 
ideal gas.  Nevertheless, we find the same functional form for the 
non-equilibrium relaxation process given by the solution 
of Eq.~(\ref{eqn:dgl_x}), which confirms 
our theoretical understanding of this rectification mechanism.  
We showed that thermal fluctuations in a single heatbath can be used to 
get directed motion if 
(i)  the motor is sustained in a non-equilibrium state by stopping it at periodically spaced sites 
on a one-dimensional track (or alternatively at periodic time intervals);  
(ii) the spatial symmetry is broken by an asymmetric shape of the motor  
which makes the non-stationary relaxation process also asymmetric. 
The shape-dependence  of the non-equilibrium relaxation is used to get 
directed motion even against small fluid drifts. By changing the shape of the motor, it 
is possible to change the direction of motion and to control the average velocity of 
the motor.


\begin{thebibliography}{99}


\bibitem{einstein05} 
A. Einstein, 
% {\"Uber die von der molekularkinetischen Theorie der W\"arme geforderte 
% Bewegung von in ruhenden Fl\"ussigkeiten suspendierten Teilchen}, 
Annalen der Physik {\bf 17}, 549 (1905). 

\bibitem{smoluchowski06}
M. v. Smoluchowski, 
% {\it Zur kinetischen Theorie der Brownschen Molekularbewegung 
% und der Suspensionen}, 
Annalen der Physik {\bf 21}, 756-780 (1906).  



\bibitem{reimann:2002}
P. Reimann, Phys. Rep. {\bf 361}, 57 (2002).  

% {\bibitem{reimann-haenggi:2002}
% {P. Reimann and P. Hänggi,  
% {\it Introduction to the physics of Brownian motors},
% {Applied Physics A: Materials \& Processing {\bf 75}, 169  (2002).  

\bibitem{feynman:63}
R. Feynman, R. Leighton and M. Sands,  
{\it The Feynman-Lectures on Physics I},
(Addison-Wesley, Reading, 1963), Chap. 46. 

\bibitem{broeck-meurs-kawai:2004}
C. Van den Broeck, R. Kawai and P. Meurs,   
% {\it Microscopic Analysis of a Thermal Brownian Motor},
Phys. Rev. Lett. {\bf 93}, 090601 (2004); \   
New Journal of Physics {\bf 7}, 10  (2005).  

% \bibitem{broeck-meurs-kawai:2005}
% C. Van den Broeck, P. Meurs, and R. Kawai,  
% {\it From Maxwell demon to Brownian motor},
% New Journal of Physics {\bf 7}, 10  (2005).  


\bibitem{howard:1996}
J. Howard,  
{\it The movement of kinesin along microtubules},
Annual Review of Physiology {\bf 58}, 703 (1996).  




\bibitem{cleuren:2007}
B. Cleuren et al.,  
% B. Cleuren and C. Van den Broeck,
% {\it Granular Brownian motor},
Europhys. Lett. {\bf 77}, 50003 (2007). 

\bibitem{costantini:2007} 
G. Costantini et al., 
%  G. Costantini, U.M.B. Marconi, and A. Puglisi,
% {\it A Granular Brownian Ratchet Model},
Phys. Rev. E 75, 061124 (2007). 


\bibitem{sporer:2006}
S. Sporer,  
{\it Ein Modell f\"ur gerichtete Brownsche Bewegung in einem einzelnen W\"armebad},
diploma thesis (Erlangen, 2006). 
% Friedrich-Alexander-Universit\"at Erlangen-N\"urnberg (2006).  

% \bibitem{mecke:1994}
% K. Mecke,  
% {\it Integralgeometrie in der Statistischen Physik},
% (Verlag Harri Deutsch, Frankfurt, 1994).  



\bibitem{lecturenotes02} K. Mecke and D. Stoyan,   
{\it Morphology of Condensed Matter}, 
Lecture Notes in Physics, Vol. 600  (Springer, Berlin,  2002).  


% \bibitem{hadwiger-meier:1973}
% H. Hadwiger and C.Meier,  
% {\it Studien zur vektoriellen Integralgeometrie},
% Mathematische Nachrichten {\bf 56}, 261-268  (1973).

\end{thebibliography}
\end{document}